\PassOptionsToPackage{unicode}{hyperref}
\PassOptionsToPackage{hyphens}{url}
\PassOptionsToPackage{dvipsnames,svgnames,x11names}{xcolor}
\documentclass[
  a4paper,
]{scrartcl}
\usepackage{xcolor}
\usepackage{amsmath,amssymb}
\usepackage{iftex}
\ifPDFTeX
  \usepackage[T1]{fontenc}
  \usepackage[utf8]{inputenc}
  \usepackage{textcomp} 
\else 
  \usepackage{unicode-math} 
  \defaultfontfeatures{Scale=MatchLowercase}
  \defaultfontfeatures[\rmfamily]{Ligatures=TeX,Scale=1}
\fi
\usepackage{lmodern}
\ifPDFTeX\else
\fi
\IfFileExists{upquote.sty}{\usepackage{upquote}}{}
\IfFileExists{microtype.sty}{
  \usepackage[]{microtype}
  \UseMicrotypeSet[protrusion]{basicmath} 
}{}
\makeatletter
\@ifundefined{KOMAClassName}{
  \IfFileExists{parskip.sty}{%
    \usepackage{parskip}
  }{
    \setlength{\parindent}{0pt}
    \setlength{\parskip}{6pt plus 2pt minus 1pt}}
}{
  \KOMAoptions{parskip=half}}
\makeatother
\makeatletter
\ifx\paragraph\undefined\else
  \let\oldparagraph\paragraph
  \renewcommand{\paragraph}{
    \@ifstar
      \xxxParagraphStar
      \xxxParagraphNoStar
  }
  \newcommand{\xxxParagraphStar}[1]{\oldparagraph*{#1}\mbox{}}
  \newcommand{\xxxParagraphNoStar}[1]{\oldparagraph{#1}\mbox{}}
\fi
\ifx\subparagraph\undefined\else
  \let\oldsubparagraph\subparagraph
  \renewcommand{\subparagraph}{
    \@ifstar
      \xxxSubParagraphStar
      \xxxSubParagraphNoStar
  }
  \newcommand{\xxxSubParagraphStar}[1]{\oldsubparagraph*{#1}\mbox{}}
  \newcommand{\xxxSubParagraphNoStar}[1]{\oldsubparagraph{#1}\mbox{}}
\fi
\makeatother

\usepackage{longtable,booktabs,array}
\usepackage{calc} 
\usepackage{etoolbox}
\makeatletter
\patchcmd\longtable{\par}{\if@noskipsec\mbox{}\fi\par}{}{}
\makeatother
\IfFileExists{footnotehyper.sty}{\usepackage{footnotehyper}}{\usepackage{footnote}}
\makesavenoteenv{longtable}
\usepackage{graphicx}
\makeatletter
\newsavebox\pandoc@box
\newcommand*\pandocbounded[1]{
  \sbox\pandoc@box{#1}%
  \Gscale@div\@tempa{\textheight}{\dimexpr\ht\pandoc@box+\dp\pandoc@box\relax}%
  \Gscale@div\@tempb{\linewidth}{\wd\pandoc@box}%
  \ifdim\@tempb\p@<\@tempa\p@\let\@tempa\@tempb\fi
  \ifdim\@tempa\p@<\p@\scalebox{\@tempa}{\usebox\pandoc@box}%
  \else\usebox{\pandoc@box}%
  \fi%
}
\def\fps@figure{htbp}
\makeatother

\NewDocumentCommand\citeproctext{}{}
\NewDocumentCommand\citeproc{mm}{%
  \begingroup\def\citeproctext{#2}\cite{#1}\endgroup}
\makeatletter
 \let\@cite@ofmt\@firstofone
 \def\@biblabel#1{}
 \def\@cite#1#2{{#1\if@tempswa , #2\fi}}
\makeatother
\newlength{\cslhangindent}
\newlength{\csllabelwidth}
\newenvironment{CSLReferences}[2] 
 {\begin{list}{}{%
  \setlength{\itemindent}{0pt}
  \setlength{\leftmargin}{0pt}
  \setlength{\parsep}{0pt}
  \ifodd #1
   \setlength{\leftmargin}{\cslhangindent}
   \setlength{\itemindent}{-1\cslhangindent}
  \fi
  \setlength{\itemsep}{#2\baselineskip}}}
 {\end{list}}
\usepackage{calc}

\providecommand{\tightlist}{%
  \setlength{\itemsep}{0pt}\setlength{\parskip}{0pt}}

\usepackage[noblocks]{authblk}

\makeatletter
\@ifpackageloaded{caption}{}{\usepackage{caption}}
\AtBeginDocument{%
\ifdefined\contentsname
  \renewcommand*\contentsname{Table of contents}
\else
  \newcommand\contentsname{Table of contents}
\fi
\ifdefined\listfigurename
  \renewcommand*\listfigurename{List of Figures}
\else
  \newcommand\listfigurename{List of Figures}
\fi
\ifdefined\listtablename
  \renewcommand*\listtablename{List of Tables}
\else
  \newcommand\listtablename{List of Tables}
\fi
\ifdefined\figurename
  \renewcommand*\figurename{Figure}
\else
  \newcommand\figurename{Figure}
\fi
\ifdefined\tablename
  \renewcommand*\tablename{Table}
\else
  \newcommand\tablename{Table}
\fi
}
\@ifpackageloaded{float}{}{\usepackage{float}}
\floatstyle{ruled}
\@ifundefined{c@chapter}{\newfloat{codelisting}{h}{lop}}{\newfloat{codelisting}{h}{lop}[chapter]}
\floatname{codelisting}{Code Block}

\makeatother
\makeatletter
\@ifpackageloaded{caption}{}{\usepackage{caption}}
\@ifpackageloaded{subcaption}{}{\usepackage{subcaption}}
\makeatother
\usepackage{bookmark}
\IfFileExists{xurl.sty}{\usepackage{xurl}}{} 
\hypersetup{
  pdftitle={In Pursuit of Total Reproducibility},
  pdfauthor={Moritz E. Beber},
  pdfkeywords={computational systems biology, metabolic modeling, event
sourcing, event modeling, reproducibility},
  colorlinks=true,
  linkcolor={blue},
  filecolor={Maroon},
  citecolor={Blue},
  urlcolor={Blue},
  pdfcreator={LaTeX via pandoc}}

\title{In Pursuit of Total Reproducibility}

\subtitle{Marrying Event Sourcing with Computational Systems Biology}

  \author{Moritz E. Beber}
            \affil{%
                  Institute for Globally Distributed Open Research and
                  Education (IGDORE)
              }
      
\date{}
\begin{document}
\maketitle
\begin{abstract}
The vast majority of scientific contributions in the field of
computational systems biology are based on mathematical models. These
models can be broadly classified as either dynamic (kinetic) models or
steady-state (constraint-based) models. They are often described in
specific markup languages whose purpose is to aid in the distribution
and standardization of models. Despite numerous established standards in
the field, reproducibility remains problematic due to the substantial
effort required for compliance, diversity of implementations, and the
lack of proportionate rewards for researchers. This article explores the
application of event sourcing - a software engineering technique where
system state is derived from sequential recorded events - to address
reproducibility challenges in computational systems biology. Event
sourcing, exemplified by systems like git, offers a promising solution
by maintaining complete, immutable records of all changes to a model.
Through examples including leader and follower applications, local and
remote computation, and contribution tracking, this work demonstrates
how event-sourced systems can automate standards compliance, provide
comprehensive audit trails, enable perfect replication of processes,
facilitate collaboration, and generate multiple specialized read models
from a single event log. An implementation of the outlined principles
has the potential to transform computational systems biology by
providing unprecedented transparency, reproducibility, and collaborative
capabilities, ultimately accelerating research through more effective
model reuse and integration. An event-sourced approach to modeling in
computational systems biology may act as an example to related
disciplines and contribute to ending the reproducibility crisis plaguing
multiple major fields of science.
\end{abstract}

\subsection{Introduction}\label{sec-intro}

\subsubsection{Reproducibility Crisis}\label{reproducibility-crisis}

Although the field of computational systems biology is, in principle, in
the enviable situation to be able to completely and deterministically
define its own tools and artefacts, like many other sciences, it has
been caught in the reproducibility crisis
(\citeproc{ref-baker_1500_2016}{Baker 2016};
\citeproc{ref-ioannidis_why_2005}{John P. A. Ioannidis 2005}).
Reproducibility, the ability of independent parties to replicate results
by merely using their methodological description
(\citeproc{ref-blinov_practical_2021}{Blinov et al. 2021};
\citeproc{ref-bacon_opus_2016}{Bacon and Burke 2016}), is integral to
the scientific method (\citeproc{ref-john_scientific_2017}{John 2017}).
Reproducibility is not only vital to the house of cards of scientific
progress, but also to the credibility of systems biology models in the
eyes of the wider public, as well as their eligibility for use in
critical decision making (\citeproc{ref-tatka_adapting_2023}{Tatka et
al. 2023}; \citeproc{ref-d_waltemath_how_2016}{D. Waltemath and O.
Wolkenhauer 2016}).

As reviewed by Blinov et al.
(\citeproc{ref-blinov_practical_2021}{2021}) and D. Waltemath and O.
Wolkenhauer (\citeproc{ref-d_waltemath_how_2016}{2016}), reproducibility
has been called into question in the related fields of computational
biology, bioinformatics, medicine, artificial intelligence,
neuroscience, and directly in computational systems biology by Tiwari et
al. (\citeproc{ref-tiwari_reproducibility_2021}{2021}). On one hand,
many of these are computational fields, so the lack of reproducibility
is surprising. On the other hand, the nature of our scientific progress
is such that we are building ever more complex models. As remarked by
Blinov et al. (\citeproc{ref-blinov_practical_2021}{2021}), combining
credible systems biology models from various sources is a matter of
accelerating research. Starting every modeling project from scratch is
simply not feasible within our short life spans and even shorter period
of graduate studies.

\subsubsection{Standards}\label{standards}

There exist ample standards in computational systems biology
(\citeproc{ref-tatka_adapting_2023}{Tatka et al. 2023};
\citeproc{ref-niarakis_addressing_2022}{Niarakis et al. 2022};
\citeproc{ref-blinov_practical_2021}{Blinov et al. 2021};
\citeproc{ref-d_waltemath_how_2016}{D. Waltemath and O. Wolkenhauer
2016}) defined by and for the community, that, when consistently
applied, facilitate reuse, interoperability, and reproducibility of
models. Although sharing research data and reproducible models is
correlated with more citations
(\citeproc{ref-hopfl_bayesian_2023}{Höpfl, Pleiss, and Radde 2023};
\citeproc{ref-piwowar_sharing_2007}{Piwowar, Day, and Fridsma 2007}),
the substantial effort required to conform with all the standards, is
rewarded too little (\citeproc{ref-ioannidis_increasing_2014}{John P. A.
Ioannidis et al. 2014}). A situation that remains unchanged despite the
clear benefits at an institutional level
(\citeproc{ref-sandve_ten_2013}{Sandve et al. 2013}).

The main standards to consider are CellML
(\citeproc{ref-clerx_cellml_2020}{Clerx et al. 2020}), NeuroML
(\citeproc{ref-gleeson_neuroml_2010}{Gleeson et al. 2010}), or SBML
(\citeproc{ref-hucka_systems_2019}{Hucka et al. 2019}) for model
definition. BioPAX (\citeproc{ref-demir_biopax_2010}{Demir et al. 2010})
for describing biological pathways. MIRIAM
(\citeproc{ref-novere_minimum_2005}{Novère et al. 2005}) for the
critical task of component annotation, which requires reference to many
resources. SBO (\citeproc{ref-courtot_controlled_2011}{Courtot et al.
2011}; \citeproc{ref-juty_systems_2010}{Juty 2010}) for anchoring the
intended meaning of model components. PEtab
(\citeproc{ref-schmiester_petabinteroperable_2021}{Schmiester et al.
2021}) for describing parameter estimation. MIASE
(\citeproc{ref-waltemath_minimum_2011}{Waltemath, Adams, Beard, et al.
2011}) and SED-ML (\citeproc{ref-waltemath_reproducible_2011}{Waltemath,
Adams, Bergmann, et al. 2011}) to describe model simulations. KiSAO
(\citeproc{ref-zhukova_kinetic_2011}{Zhukova et al. 2011}) for
describing simulation algorithms used. Finally, OMEX to describe
metadata and package everything into a COMBINE archive
(\citeproc{ref-bergmann_combine_2014}{Bergmann et al. 2014}). As even
the uninitiated can imagine, following all of the above standards to the
letter, especially manually annotating all model components, requires an
amount of effort that runs counter to today's fast publishing climate
and poses a challenge even to experts.

\subsubsection{Marrying Event Sourcing}\label{marrying-event-sourcing}

Event sourcing is a software engineering technique that is gaining in
popularity, but whose origins are somewhat unclear. Suffice to say that
it relates to modern double-entry book keeping\footnote{\href{https://en.wikipedia.org/wiki/Double-entry_bookkeeping\#History}{According
  to Wikipedia}, the practice started around the 13th century in Europe.
  Probably much earlier in other parts of the world {[}author's note{]}.}
and has lately been popularized in a series of talks by Greg
Young\footnote{The technique was possibly first published in rudimentary
  form by Martin Fowler
  \href{https://martinfowler.com/eaaDev/EventSourcing.html}{on his blog}
  in 2005.} \footnote{See, for example,
  \href{https://www.youtube.com/watch?v=8JKjvY4etTY}{this recording at
  GOTO 2014}. A book on the topic is apparently in preparation.}. The
core idea of event sourcing is that application state is derived from
all past events which are records of facts; just like your bank
account's balance is the sum of all transactions. Besides banking, a
prime example of event sourcing is \href{https://git-scm.com/}{git}. A
version control software used by the vast majority of the software
industry. In this work, I explain event sourcing for computational
systems biologists and outline several applications of this technique
for the task of automating generating standards compliant models,
provide critical infrastructure, foster collaboration, and subsequently
improve reproducibility.

In trying to design and understand event-sourced systems, there have
been numerous influences. Early on, Domain-Driven Design's (DDD -
\citeproc{ref-evans_domain-driven_2004}{Evans 2004}) aggregates and the
ubiquitous language were a large inspiration. The ubiquitous language
means that all software artifacts, like documentation, code, and tests
are using the same concepts and language as the domain being modelled.
Aggregates are objects that ensure the consistency of their data within
their context. That means, they contain rules that may disallow commands
issued to the aggregate from succeeding and thus control the state of
the aggregate. Greg Young's work on Command Query Responsibility
Segregation (CQRS) was key to more functional implementations of event
sourcing and led to the realization that it is not only possible, but in
fact desirable to have multiple specialized read models to serve
information queries. Designing event-sourced systems was helped by a
workshop format called Event Storming by Alberto Brandolini\footnote{https://www.eventstorming.com}.
Today, many practitioners follow Adam Dymitruk's Event
Modeling\footnote{https://eventmodeling.org}, which was recently more
formally published by Dilger
(\citeproc{ref-dilger_understanding_2024}{2024}).

\subsection{Methods}\label{methods}

\subsubsection{Event Sourcing in a Nut
Shell}\label{event-sourcing-in-a-nut-shell}

Most software applications, such as those used for computational
simulations of biological systems, store the current state, for example,
the state of a model, its parameter values, and so on, in memory. In
order to reuse that current state, and avoid complete loss of all
information stored in memory at shutdown, most applications allow for
storing the state in a file or database. That means that we might make
many modifications to the application state, but only ever store a
representation of the latest state. With event sourcing, however, the
application state is derived from sequential events. It is that sequence
of events that is then stored in a more permanent fashion in a file or
database; and the application state can be restored by reading the
series of events and applying their recorded information.

What are the benefits of storing all sequential records of change?
Throughout this work, I will attempt to convince you of several
advantages of event sourcing, most importantly, the improved
reproducibility afforded by this technique.

\paragraph{Liquid Handling Example}\label{liquid-handling-example}

Let us consider the following scenario, we need to generate a growth
medium for our cell cultures. We are in possession of a protocol that
lists in minute detail which materials to combine in what quantities in
order to arrive at the final growth medium in exact proportions. When we
follow the protocol, we can assume that we arrive at the correct medium.
However, the only way to confirm that fact, is to perform an intricate
chemical analysis of our solution. The receptacle with our chemical
solution represents our ``application state''.

\begin{figure}

\centering{

\pandocbounded{\includegraphics[keepaspectratio]{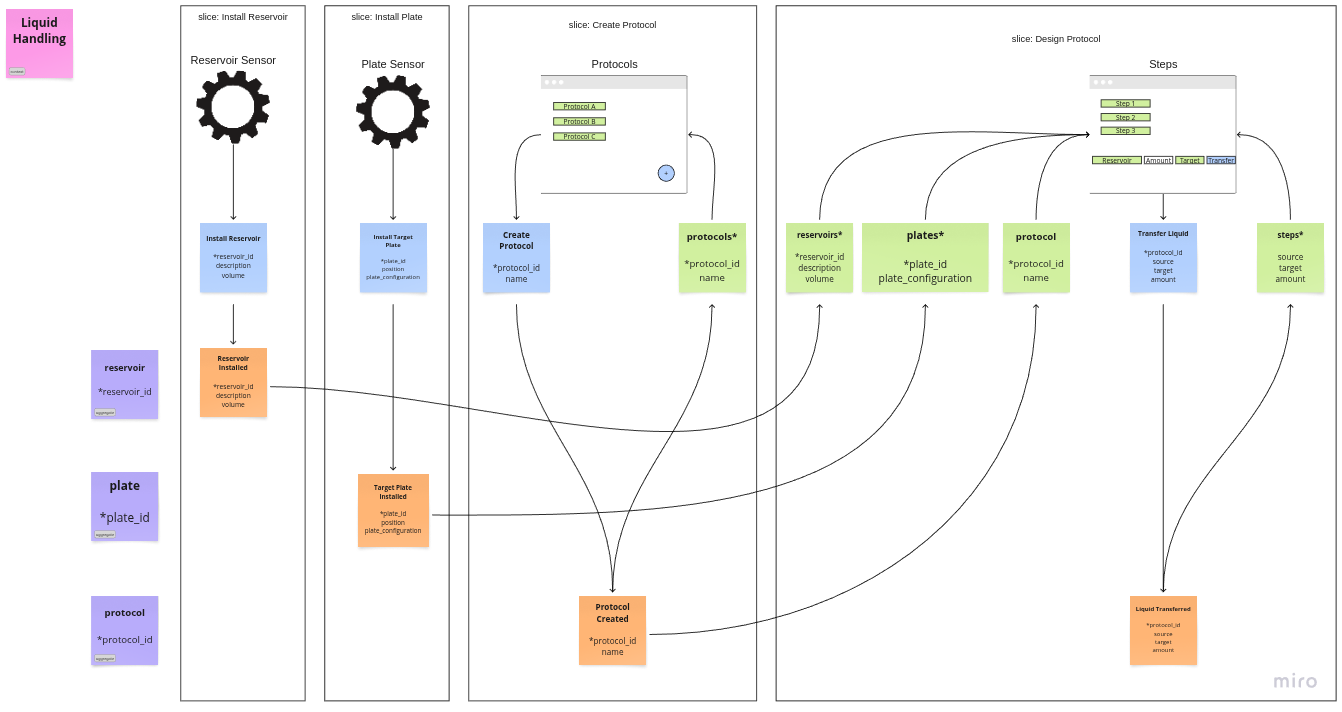}}

}

\caption{\label{fig-liquid-handling}A simplified event model showing the
addition of reservoirs, target well plates, protocols, and liquid
transfers. Commands are shown in light blue, events are orange, and read
models are light green.}

\end{figure}

Let us then consider a hypothetical, vastly simplified, liquid handling
robot with an event-sourced operating software. The event model
(Figure~\ref{fig-liquid-handling}), clearly shows that we can add
reservoirs with basic components of our medium to the robot and we can
load a 96-well plate (0.2 mL volume per well) where to perform the
mixing. We can also see a list of protocols and create a new protocol.
Within a single protocol, we can see the individual steps and add
further liquid transfers as new steps.

\begin{figure}

\centering{

\pandocbounded{\includegraphics[keepaspectratio]{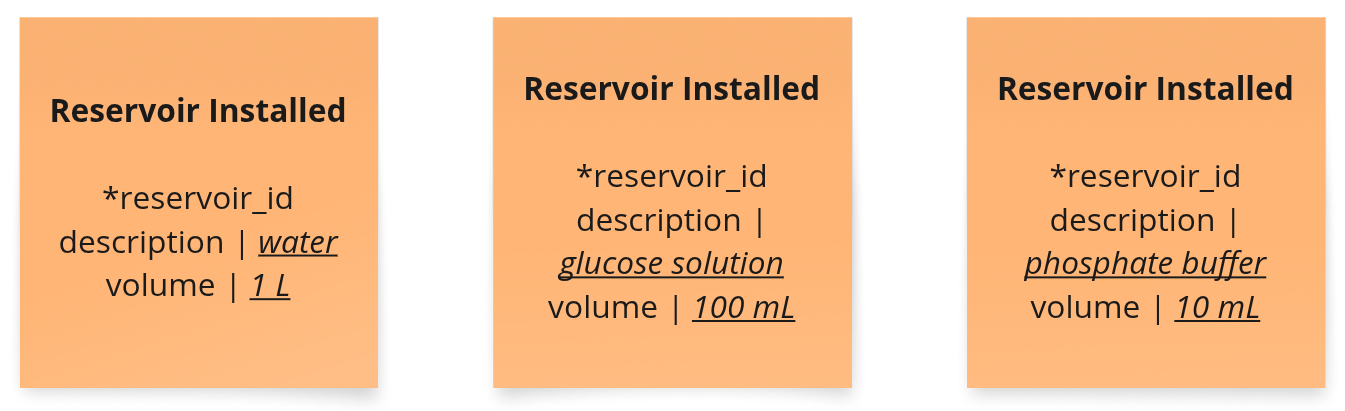}}

}

\caption{\label{fig-reservoirs-installed}Three sequential events that
record information about reservoirs that were installed in the robot.}

\end{figure}

Assuming that adding reservoirs with water, glucose solution, and
phosphate buffer as sources will automatically be detected by a sensor,
three commands are issued by the sensor and assuming that there are no
errors, three events record those facts (see
Figure~\ref{fig-reservoirs-installed}). Additionally, we install a
96-well plate and another sensor identifies its configuration and issues
a corresponding command. Assuming the command succeeds, a \textbf{Target
Plate Installed} event is recorded.

What can we already learn from the few events shown in
Figure~\ref{fig-reservoirs-installed}?

\begin{enumerate}
\def\labelenumi{\arabic{enumi}.}
\tightlist
\item
  By convention, the names of events are chosen in past tense, because
  they are records of information.
\item
  There is a global order to the events as denoted here by the
  left-to-right arrangement.
\end{enumerate}

Our protocol for the growth medium might ask for 90\% water, 9\% glucose
solution, and 1\% phosphate buffer. We might therefore issue the
following instructions. Using the protocols interface, we first create a
new protocol with the name, ``Minimal Glucose Growth Medium''. We then
define three steps for our protocol:

\begin{enumerate}
\def\labelenumi{\arabic{enumi}.}
\tightlist
\item
  Transfer 180 \(\mu\)L of water from its reservoir to the target plate.
\item
  Transfer 18 \(\mu\)L of glucose solution from its reservoir to the
  target plate.
\item
  Transfer 2 \(\mu\)L of phosphate buffer from its reservoir to the
  target plate.
\end{enumerate}

When these commands are issued, the operating software will use the
information present to perform a range of consistency checks. Does the
reservoir hold enough liquid for the requested transfer? Is there enough
volume left in the target plate to accept the liquid transfer? There
might be further rules as to which chemicals are safe to combine, and so
on. Assuming that all consistency checks pass, three \textbf{Liquid
Transferred} events are recorded. We could then execute our protocol and
let the robot prepare our growth medium.

What can we learn from this train of thoughts in addition to what we
have already deduced?

\begin{enumerate}
\def\labelenumi{\arabic{enumi}.}
\tightlist
\item
  Events only store specific properties related to what is being
  recorded.
\item
  The sequence of events provides a detailed log of everything that
  occurred in the system.
\end{enumerate}

All the events together form what is called the \textbf{event log} and
it is of central importance. If we want to know (query) the state of an
instance at a certain point in the sequence of events, we need to apply
the information recorded in all events concerning that instance in order
to derive the state. This process is called a \textbf{projection}. We
have seen that events can be quite verbose. Fortunately, computational
storage space is generally fairly cheap.

After issuing all commands and executing our protocol, the resulting
growth medium in the wells is the same as if we followed the
experimental protocol manually. That means, the ``real world'' state is
the same as before, however, the event log enables us to do more.

\begin{enumerate}
\def\labelenumi{\arabic{enumi}.}
\tightlist
\item
  We can analyze the sequence of events to show that we have correctly
  implemented the experimental protocol. This is much more convenient
  than having to perform chemical analyses on the resulting medium.
\item
  We can investigate the sequence of events at any point that we desire
  in order to identify potential mistakes that we may have made. This is
  also called \textbf{replaying} events.
\item
  Similarly, we may look for optimized sequences of events. This may be
  done by simulating the operation of the liquid handling robot and
  exploring alternative orders of commands and events.
\item
  We can copy our event log and hand it to somebody else. That other
  person can then perfectly replicate our process by replaying the
  events and arriving at the same outcome. This serves the same purpose
  as a published experimental protocol but provides much stronger
  reproducibility guarantees as the operating software assures that we
  perform a perfect replicate of the process.
\end{enumerate}

\subsection{Results}\label{results}

In the following, I will outline a few examples that are enabled by
applying event sourcing to computational systems biology. They each
highlight specific advantages of this technique and in particular the
benefits of an event log. We already discussed general benefits in the
liquid handling example above, but I want to emphasize again how the
event log can alleviate problems observed with current publication
practices. For convenience, typically only a single SBML document
describing a particular model state is attached to a publication.
However, an analysis described in a publication needs to explore many
variations of this model. In the best case, these alternative scenarios
are available as source code files that load the SBML document and
modify the model further via code instructions. In the worst case, such
modifications are not even described in text form, thus severely
hampering reproducibility of published results.

With an event-sourced simulation software, it would be possible to
provide the full event log and specify which published result was
obtained at which point in the event log. Others can then replay the
events to that point in the log, and reproduce the model state
perfectly.

\subsubsection{Contribution History}\label{contribution-history}

Scientific contributions and public records demonstrating said
contributions, are of vital importance to academics. In the context of
computational systems biology, one major contribution is the definition
of a model for simulation. The \emph{de facto} standard for defining
such models in a machine readable format is the Systems Biology Markup
Language (SBML - \citeproc{ref-hucka_systems_2019}{Hucka et al. 2019}).
Although the SBML specification\footnote{\href{https://identifiers.org/combine.specifications:sbml.level-3.version-2.core.release-2}{SBML
  Level 3 Version 2 Core, Section 6.6, pp.~105.}} allows for recording
contributions at a very detailed level, i.e., every element in the
document may contain annotations with contributors described in the
vCard4 format\footnote{\url{https://datatracker.ietf.org/doc/html/rfc6350}},
in practice, I have only seen this done at the level of the entire
model.

It is certainly not surprising that the full amount of detail afforded
by the SBML specification is not used, as it would be very cumbersome to
manually track this information. However, if we were to create an SBML
document with an event sourced software, we could identify each user of
that software by, for example, an \href{https://orcid.org/}{ORCID}, and
reference the user in the event that recorded the information. This
could serve as an audit trail of contributions to the model. Such a
software would then require functionality to transform the event log
into an SBML document that correctly uses SBML annotations to detail the
modification history of every model element. Other tools could then
create accurate contribution statistics from such SBML documents. This
example already leads us to the next section, since creating an SBML
document from an event log is nothing else but a projection or
alternative read model.

\subsubsection{Alternative Read Models}\label{alternative-read-models}

\begin{figure}

\centering{

\pandocbounded{\includegraphics[keepaspectratio]{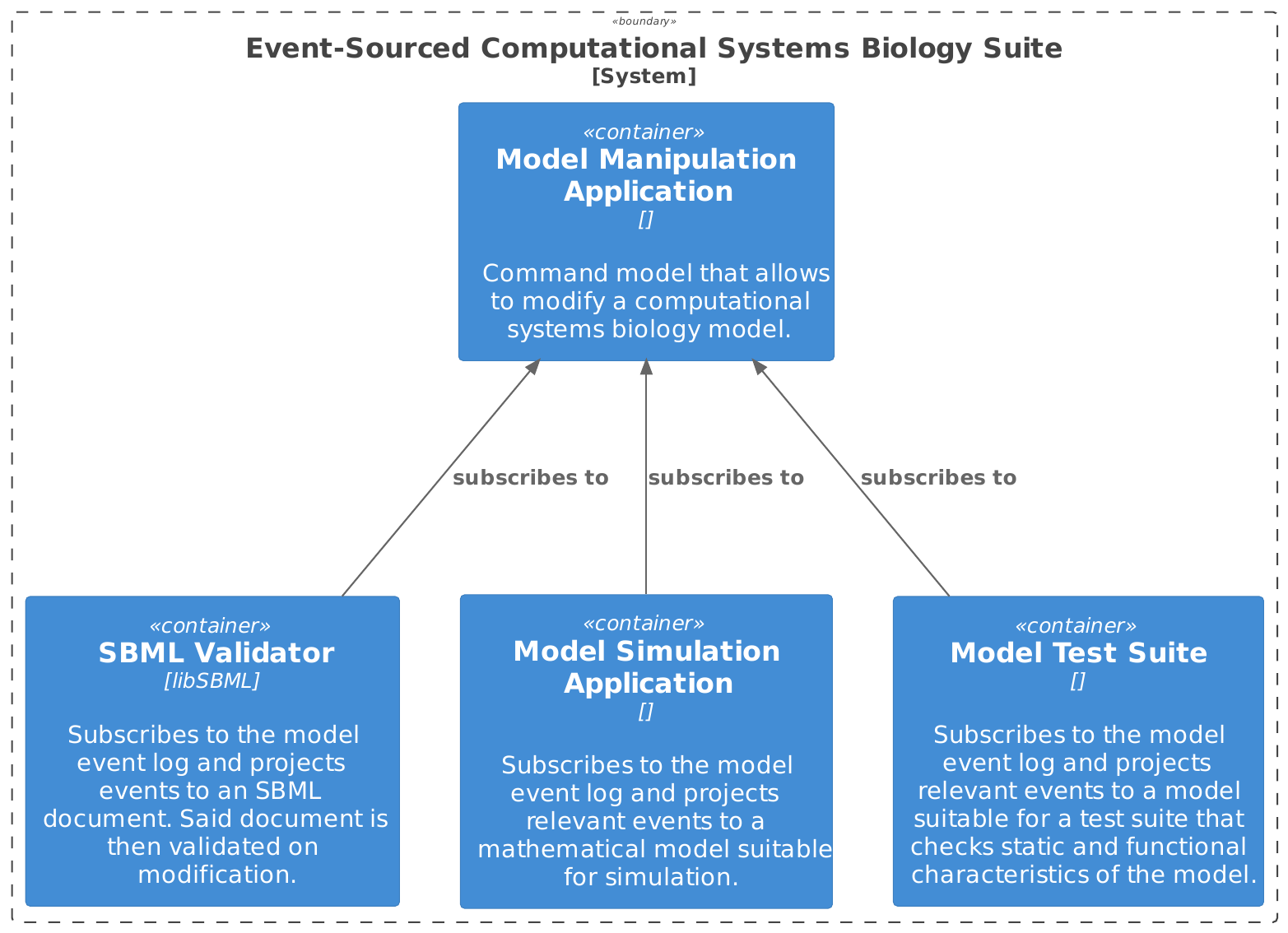}}

}

\caption{\label{fig-alternative-models}A system of event-sourced
applications with one leader and three followers.}

\end{figure}

Let us assume that we have an event-sourced software for manipulating
computational systems biology models. Most often, those will be either
kinetic or constraint-based models. Our software will contain a
\textbf{model manipulation application}. Our application will produce an
event log that can then be read by other applications. Those other
applications can then project those events into other formats.
Importantly, those formats may only be used for reading, i.e., they are
query models, whereas modifications must be made through commands on the
previous model manipulation application. Our event-sourced software may
thus consist of a system of applications with several downstream query
models constructed from following one or more upstream event logs. An
example of such a system is shown in
Figure~\ref{fig-alternative-models}.

In Figure~\ref{fig-alternative-models}, we see three examples of
downstream applications that subscribe to the event log of the model
manipulation application. It is important to note that each downstream
application will create different projections of those events. The shown
SBML validator will probably make use of every detail in every event,
meticulously mapping all content to an SBML document and validating
that. A model simulation application for, e.g., a constraint-based
metabolic model will likely only be interested in events describing the
stoichiometry of reactions and metabolites, projecting that information
into (linear) constraints of a mathematical optimization problem. A
fictitious model test suite will use yet another set of events depending
on the requirements of its checks. It might even include a model
simulation application for functional tests.

\subsubsection{Local Versus Remote}\label{local-versus-remote}

\begin{figure}

\centering{

\pandocbounded{\includegraphics[keepaspectratio]{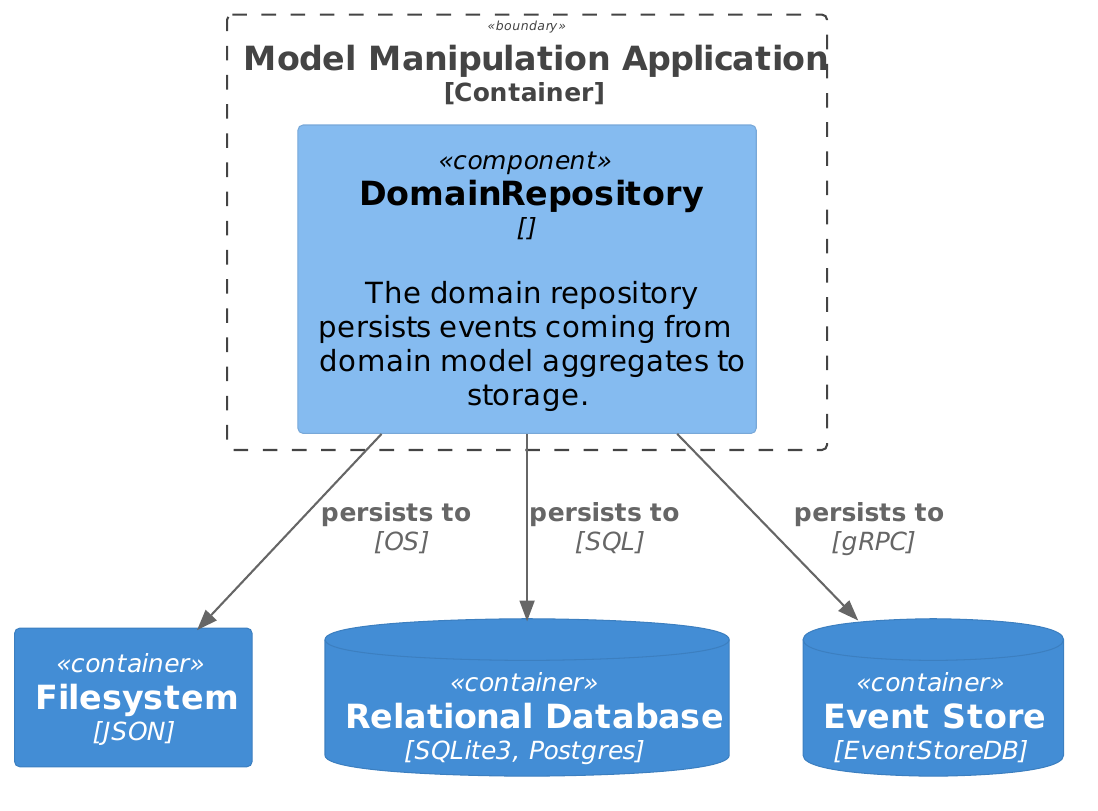}}

}

\caption{\label{fig-event-log-persistence}Persistence of events may
happen in different locations. On the local filesystem; in local or
remote databases; or in cloud hosted event stores.}

\end{figure}

Another important benefit of event-driven software architectures is that
they afford asynchronicity and decoupling in systems. Asynchronicity
because a system only needs to react to events when they arrive, and
decoupling because upstream applications need not know anything about
the downstream subscribers to their event logs. For these reasons, it is
a smaller change to switch from a local-only to a remote implementation.
A downside of events is increased latency due to message passing. An
example of a seamless change between a local and a remote implementation
is event persistence as shown in Figure~\ref{fig-event-log-persistence}.
A remote database may be hosted anywhere in the cloud, but even our
different applications from Figure~\ref{fig-alternative-models} might be
globally distributed by using an appropriate event bus.

\begin{figure}

\centering{

\pandocbounded{\includegraphics[keepaspectratio]{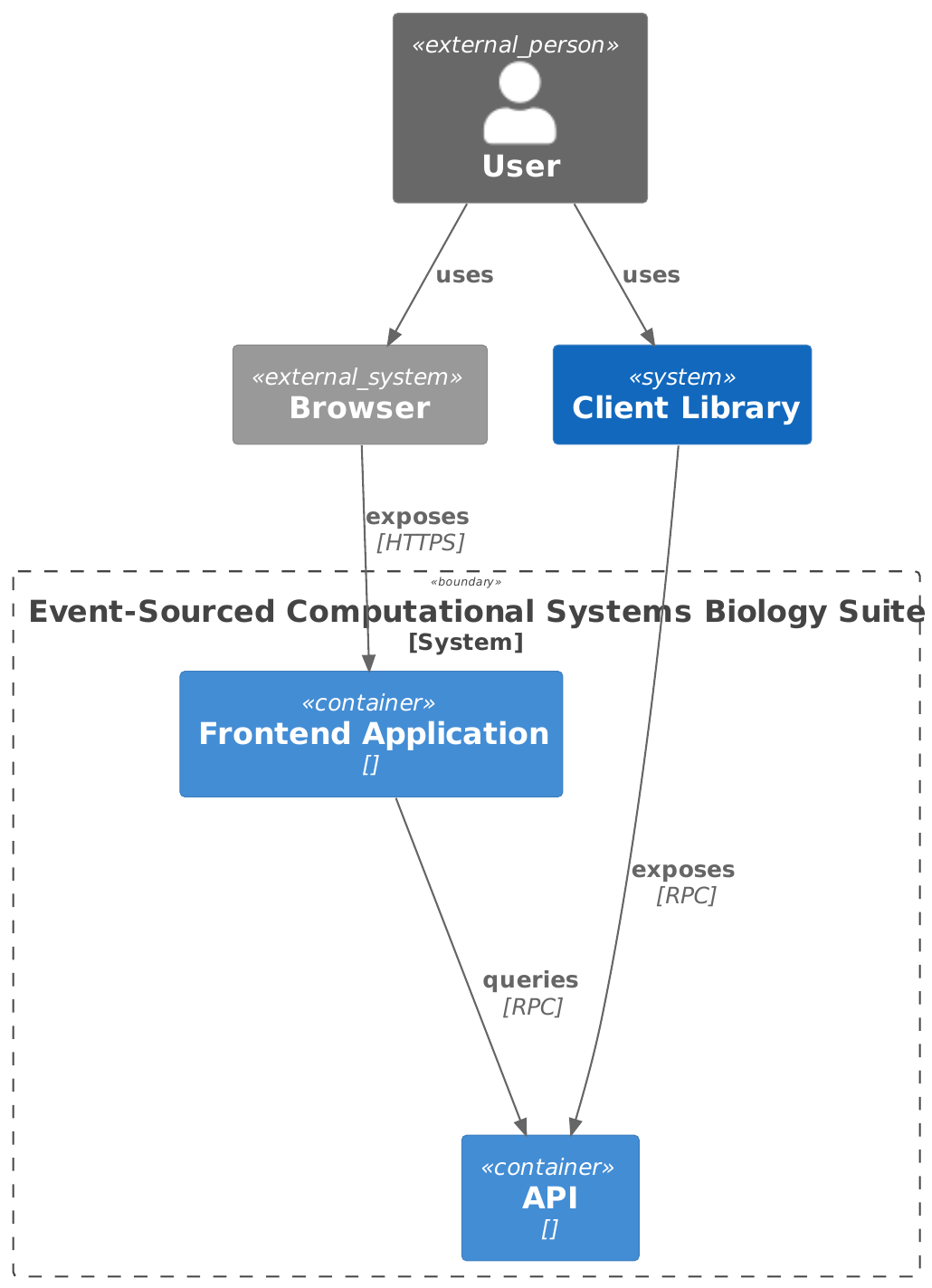}}

}

\caption{\label{fig-escsbs}A depiction of a hypothetical system that
exposes a frontend application to browsers, as well as a client library
that directly communicates with a backend API.}

\end{figure}

We can take this further by considering that a user of this software may
not even need to have installed our software system, but can interact
with the model manipulation and other applications via a client that
performs \textbf{remote procedure calls} (RPCs). This client might
either be a graphical interface or a small library of functions that
exposes the \textbf{application programming interface} (API) of our
system. These two modes of interaction are visualized in
Figure~\ref{fig-escsbs}. Besides circumventing installation, this design
has two additional benefits:

\begin{enumerate}
\def\labelenumi{\arabic{enumi}.}
\tightlist
\item
  Mathematical solvers for a model simulation application may be
  proprietary or particularly difficult to install. In this way, we can
  offer these capabilities to a wider audience. Something similar was
  done by Shaikh et al.
  (\citeproc{ref-shaikh_biosimulators_2022}{2022}).
\item
  It is less work to develop and maintain a client library compared to
  the full system. Thus we can reach a larger user base by providing
  clients in multiple programming languages.
\end{enumerate}

A popular example of such a setup is
\href{https://jupyter.org/}{Jupyter}
(\citeproc{ref-b_e_granger_jupyter_2021}{B. E. Granger and F. Pérez
2021}), which offers a rich Python, Julia, or R interpreter via a web
interface. The concept of managed sessions that users can connect to
from any device is also extremely interesting in this context.

\subsubsection{Collaboration Platform}\label{collaboration-platform}

As mentioned in the introduction, one of the premier examples of
event-sourced software is \href{https://git-scm.com/}{git}. ``Git is a
free and open source distributed version control system {[}\ldots{]}''.
Distributed means that git affords both local and remote work between
multiple parties. The central concept of git is that a user can record
changes to files under git's control. Each recorded set of changes is an
event, called commit. Git allows for a diverging event log, such a
diversion is called a branch. Git also has facilities for merging
branches back together and resolving conflicts between the branches.

If we take all of the above application examples together, we might
arrive at the concept of an online collaboration platform, just as, for
example, \href{https://github.com/}{GitHub} and
\href{https://gitlab.com/}{GitLab} have done for git. We already
suggested using git as part of the workflow to develop metabolic models
(in SBML) in Lieven et al. (\citeproc{ref-lieven_memote_2020}{2020}),
but here we can develop these concepts further into a full service
platform.

Such a full service online collaboration platform has many fine details
to decide on, but in broad strokes it might provide the following
features:

\begin{itemize}
\tightlist
\item
  As outlined above, such a platform should offer both graphical and
  programmatic interfaces to its API.
\item
  Every user contribution should be tracked and exposed.
\item
  There should be central repositories for all model components of
  interest that users can collaborate on improving and expanding. For
  metabolic models, for example, the definition of metabolites and
  reaction stoichiometries is of central importance.
\item
  Every process of collaboration should be backed by branching and
  merging facilities as seen in git. All suggested changes need to be
  reviewed, perhaps through processes as seen in
  \href{https://www.wikipedia.org/}{Wikipedia} and pull/merge requests
  on GitHub/GitLab.
\item
  Users would then be able to include components from the central
  repositories to develop specific models.
\item
  Every contributed change might automatically undergo a series of
  static and functional checks, allowing all collaborators to assess the
  impact of that change.
\item
  Part of the functional checks could be a comparison of simulated
  predictions with real world data.
\item
  Each version of a model, given by its history of changes, will
  automatically be available as an SBML document.
\item
  The exact contributors to a model and the size of their contributions
  should be publicly visible and considered on publication.
\end{itemize}

\subsection{Discussion}\label{discussion}

In this work, I have introduced the event sourcing technique to
computational systems biology and outlined several potential
applications in this field. I am convinced that the complete history of
changes to models provided by their event logs, as well as the ability
to exchange and perfectly reproduce processes via the event log, can
overcome most of today's challenges with reproducibility. Furthermore,
the new possibilities for collaboration and transparency have the
potential to transform and accelerate how we work in computational
systems biology today. By cleverly chaining applications that follow
each others' event logs, standards-compliant outputs could be produced
automatically, avoiding countless hours of manual labor as well as
mistakes.

Although event sourcing, as a technique that represents a paradigm
shift, certainly presents complications, I see the main challenge for
bringing its benefits to bear in finding the right funding that can
attract the skilled experts required to build and support a large
software ecosystem.

\subsection{Acknowledgements}\label{acknowledgements}

I would like to thank John Bywater, the author of a
\href{https://eventsourcing.readthedocs.io/}{Python framework for event
sourcing}, for his patience in answering my questions and his openness
to general discussions. I would also like to thank
\href{https://orcid.org/0000-0003-1238-1499}{Zachary A. King, PhD} for
showing me an early prototype of ``lifelike'', making me think about
online collaboration for metabolic models.

\section*{References}\label{bibliography}
\addcontentsline{toc}{section}{References}

\phantomsection\label{refs}

\end{document}